\documentclass{article}
\usepackage[a4paper,
            left=1cm,
            right=1cm,
            top=2.5cm,
            bottom=2.5cm,
            headheight=17pt]{geometry}
\usepackage{tikz}
\usetikzlibrary{arrows.meta, calc}
\usepackage{amsfonts}
\usepackage{pgfplots}
\usepgfplotslibrary{polar}
\pgfplotsset{compat=1.18}
\usetikzlibrary{decorations.markings}
\usepackage{amssymb}
\usepackage{amsmath}
\usepackage{amsthm}
\usepackage[english]{babel}
\usepackage{epigraph}
\usepackage[utf8]{inputenc} 
\usepackage[T2A]{fontenc}

\newcommand{\al}{\alpha}
\newcommand{\be}{\beta}

\newcommand{\pa}{\partial}

\newcommand{\Ga}{\Gamma}
\newcommand{\om}{\omega}

\newcommand{\ga}{\gamma}
\newcommand{\beq}{\begin{equation}}
\newcommand{\ee}{\end{equation}}
\newcommand{\bmat}{\begin{pmatrix}}
\newcommand{\emat}{\end{pmatrix}}
\newcommand{\bseq}{\begin{subequations}}
\newcommand{\eseq}{\end{subequations}}
\begin{document}

\title{Classical tests of New General Relativity}

\author{V. P. Vandeev$^1$, V. A. Parkhomenko$^2$, A. N. Semenova$^{1,2}$, M. A. Matveev$^1$\\
{\small \it $^1$Petersburg Nuclear Physics Institute of National Research Centre ``Kurchatov Institute'', \small \it Gatchina 188300, Russia}\\
{\small \it $^2$ Faculty of Physics, St. Petersburg State University, \small \it Saint Petersburg 198504, Russia}\\ 
{\small  vandeev{\_}vp@pnpi.nrcki.ru}\hspace{10 pt} \small parhomenko2.va@gmail.com \hspace{10 pt} \small ala.semenova@gmail.com\hspace{10 pt} \small matveev{\_}ma@pnpi.nrcki.ru}
\date{}

\maketitle

\begin{abstract}

In New General Relativity (NGR), the static spherically symmetric vacuum solution is characterized by two parameters, unlike General Relativity (GR), where the Schwarzschild solution has only one parameter. In this work, using the five classical tests of relativistic gravity (perihelion precession, light bending, Shapiro time delay and de Sitter geodetic precession effect, as well as gravitational redshift), a more stringent constraint on the value of the second parameter was derived compared to the original work\cite{NewGR}.

\end{abstract}

\section{Introduction}

Usually modified theories of gravity are developed to describe cosmological phenomena on large spatial scales, where General Relativity requires for such description to introduce dark energy-momentum sectors, for example to explain observed galactic rotation curves. Anoter purpose is to construct a renormalizable quantum theory of gravity. This paper will be devoted only to classical gravitational phenomena in one of the modified theories. 

One of the approaches to modify Einstein's General Relativity (GR) is teleparallelism \cite{TGin}, \cite{TGftc}, where the spacetime manifold is endowed with a Weitzenb\"{o}ck connection. Unlike the Levi-Civita connection, this connection possesses torsion but no curvature, so the formulation of a gravitational theory in the tetrad formalism called the Teleparallel Equivalent of General Relativity (TEGR) is dynamically fully equivalent to GR

In this work, we will consider the simplest generalization of TEGR, the so-called New General Relativity (NGR), where three arbitrary constants arise, whose values will be estimated using classical gravitational tests. Simplicity of NGR is based on the fact that all torsion scalars in the action are even and the equations of motion contain no derivatives higher than second order. This approach allows to modify not General Relativity (or the Einstein-Hilbert action) directly, as theories like $f(R)$ \cite{fR} or Gauss-Bonnet gravity \cite{GBG}, but its teleparallel equivalent based on non-Riemannian geometry.

Studies investigating consistency of modified gravity theories and classical Einsteinian tests are quite widely represented in modern scientific literature. See for example: \cite{fRct} about $f(R)$ gravity, \cite{bgct} about Brane gravity, \cite{tgct} on some teleparallel theories, \cite{stct} about scalar-tensor gravity. It would also be interesting to study the physical effects in axially symmetric gravitational fields in NGR and compare them with the predictions of GR and experimental data.

The paper has the following structure: Section 2 presents the fundamental facts and equations of NGR. Section 3 describes obtaining an exact static spherically symmetric vacuum solution. Section 4 is devoted to the analysis of invariants of the obtained solution. Section 5 consists of five subsections, each dedicated to a specific gravitational phenomenon: perihelion precession, light bending, Shapiro time delay, de Sitter geodetic precession and gravitational redshift; and the comparison of its observable quantity with the analogous one in GR (TEGR). Section 5 provides a summary and outlines potential future research directions for NGR and its applications.

\section{New General Relativity}
On a smooth spacetime manifold with a metric $g_{\mu\nu}$, one can locally introduce a tetrad field $e^a_\mu$ which is determined by the metric
\beq
g_{\mu\nu}=e_\mu^{a} e_\nu^{b}\eta_{ab},  
\ee
where $\eta_{ab}=\text{diag}(1,-1,-1,-1)$ is flat Minkowski metric. The covariant derivative of a tensor object with indices of two types has the form
\beq
\nabla_\mu W^{a\nu}=\partial_\mu W^{a\nu}+\Gamma^\nu_{\ \mu\rho}W^{a\rho}+\omega^a_{\ \mu c}W^{c\nu},
\ee
where $\om$ is the spin connection and $\Ga$ is the metric affine connection which defines torsion and curvature tensors
\beq
T^\sigma_{\ \mu\nu}\equiv\Gamma^\sigma_{\ \mu\nu}-\Gamma^\sigma_{\ \nu\mu},\quad R^\al_{\ \be\mu\nu}\equiv\partial_\mu\Gamma^\al_{\ \nu\be}-\partial_\nu\Gamma^\al_{\ \mu\be}+\Gamma^\al_{\ \mu\gamma}\Gamma^\gamma_{\ \nu\be}-\Gamma^\al_{\ \nu\gamma}\Gamma^\gamma_{\ \mu\be}.
\ee
The condition of teleparallelism (or absolute parallelism) is expressed as the covariant constancy of the tetrad field
\beq\label{cce}
\nabla_\mu e_{a}^\nu=\partial_\mu e_{a}^\nu+\Gamma^\nu_{\mu\al}e_{a}^\al-\omega^b_{\ \mu a}e_{b}^\nu=0,\quad
\nabla_\mu e^{a}_\nu=\partial_\mu e^{a}_\nu-\Gamma^\al_{\mu\nu}e^{a}_\al+\omega^a_{\ \mu b}e^{b}_\nu=0,
\ee
this allows one to connect the affine and spin connections
\beq
    \Gamma^\al_{\mu\nu}=e^\al_{ a}\left(\partial_\mu e_\nu^{ a}+\omega^a_{\ \mu b}e^{ b}_\nu\right),\quad
    \omega^{ a}_{\ \mu b}=-e^\nu_{ b}\partial_\mu e^{ a}_\nu+e_\al^{ a}\Gamma^\al_{\mu\nu}e^\nu_{ b}.
\ee
Therefore, for torsion and curvature, the Cartan structural equations are
\beq
T^\al_{\ \mu\nu} = e^\al_a\left(\partial_\mu e^a_{\ \nu} - \partial_\nu e^a_{\ \mu} + \omega^a_{\ b\mu} e^b_{\ \nu} - \omega^a_{\ b\nu} e^b_{\ \mu}\right),\quad R^\al_{\ \be\mu\nu} =e^\al_a e_\be^b \left(\partial_\mu \omega^a_{\ b\nu} - \partial_\nu \omega^a_{\ b\mu} + \omega^a_{\ c\mu} \omega^c_{\ b\nu} - \omega^a_{\ c\nu} \omega^c_{\ b\mu}\right).
\ee
This means that under the assumption of a trivial spin connection, the curvature becomes zero, and the connection becomes Weitzenb\"{o}ck-like \cite{WG}
\beq
    \Gamma^\al_{\mu\nu}=e^\al_{ a}\partial_\mu e_\nu^{ a},
\ee
and its torsion is the commutator of the partial derivatives of the tetrad field
\beq
T^\al_{\ \mu\nu}=e^\al_a\left(\pa_\mu e^a_{\nu}-\pa_\nu e^a_{\mu}\right).
\ee
Using this torsion tensor three independent scalars can be constructed, whose linear combination defines the Lagrangian density of the action 
  \beq\label{NGRaction}
S=\frac{1}{2\varkappa}\int dx^4|e| \left(\frac{a}{4}T^{\al\mu\nu}T_{\al\mu\nu}+\frac{b}{2}T^{\al\mu\nu}T_{\mu\al\nu}-cT^\mu T_\mu\right),
\ee
where $|e|$ is the determinant of $e^a_\mu$, and $a,b,c$ are arbitrary constants, $\varkappa$ is the gravitational coupling constant. The action (\ref{NGRaction}) for $a=b=c=1$ differs from the Einstein-Hilbert action only by a total divergence, i.e. a boundary term:
\beq
\frac{1}{4}T^{\al\mu\nu}T_{\al\mu\nu}+\frac{1}{2}T^{\al\mu\nu}T_{\mu\al\nu}-T^\mu T_\mu+2\tilde{\nabla}_\mu T^\mu=-R,
\ee
where $T_\mu=T^\al_{\ \mu\al}$ is torsion vector, $\tilde{\nabla}$ is covariant derivative with respect to the Levi-Civita connection, $R$ is Ricci scalar.
Varying action (\ref{NGRaction}) with respect to the tetrad one obtains the following vacuum equation
\beq\label{TE}
\mathfrak{G}_\mu^{\ \nu}\equiv\tilde{\nabla}_\alpha \mathfrak{S}_{\mu}^{\ \nu\alpha}-K_{\alpha\mu\beta}\mathfrak{S}^{\alpha\nu\beta}+\frac{1}{2}\mathfrak{T}\delta^{\nu}_{\mu}=0,
\ee
where $K_{\al\mu\nu}$ is the contorsion tensor, $\mathfrak{S}_{\al\mu\nu}$ is the superpotential defined as follows:
\bseq
\beq
K_{\al\mu\nu}=\frac{1}{2}\left(T_{\al\mu\nu}+T_{\mu\al\nu}+T_{\nu\al\mu}\right),
\ee
\beq
\mathfrak{S}_{\al\mu\nu}=\frac{a}{2}T_{\al\mu\nu}+\frac{b}{2}\left(T_{\mu\al\nu}-T_{\nu\al\mu}\right)+c\left(g_{\al\mu}T_\nu-g_{\al\nu}T_\mu\right).
\ee
\eseq
The generalized Einstein tensor defined in (\ref{TE}) for arbitrary $a,b,c$ generally possesses an antisymmetric part. It is useful to note that if additional matter fields are present in action (\ref{NGRaction}), the right-hand side of equation (\ref{TE}) acquires an energy-momentum tensor.
\section{Exact solution}
In a recent paper \cite{ssssngr}, using the tetrad 
\beq\label{sphtet}
e_\mu^{a}=
\begin{pmatrix}
	f & 0 & 0 & 0\\
	0 & g\sin\theta\cos\phi & g\rho\cos\theta\cos\phi & -g\rho\sin\theta\sin\phi\\
	0 & g\sin\theta\sin\phi & g\rho\cos\theta\sin\phi & g\rho\sin\theta\cos\phi\\
	0 & g\cos\theta & -g\rho\sin\theta & 0
\end{pmatrix},
\ee
it was shown that for any parameters of the theory, static spherically symmetric vacuum solutions can be expressed in elementary functions. The advantage of this tetrad is that it does not generate the antisymmetric part of the generalized Einstein tensor. It turns out that an exact vacuum solution can be found using the ansatz (\ref{sphtet}). So, the non-zero components of the Einstein equations have the form
\bseq\label{te}
\beq\label{tet}
\left[\left(d-c\right)\frac{f'}{fg^2}-2c\frac{g'}{g^3}\right]'+\left(\frac{f'}{f}+3\frac{g'}{g}+\frac{2}{\rho}\right)\left[\left(d-c\right)\frac{f'}{fg^2}-2c\frac{g'}{g^3}\right]+\frac{1}{2}\mathfrak{T}=0,
\ee
\beq\label{ter}
\left(d-c\right)\frac{f'^2}{f^2g^2}-(4c-2d)\left(\frac{g'}{g}+\frac{1}{\rho}\right)\frac{g'}{g^3}-2c\left(2\frac{g'}{g}+\frac{1}{\rho}\right)\frac{f'}{fg^2}+\frac{1}{2}\mathfrak{T}=0,
\ee
\beq\label{tea}
\left[\left(d-2c\right)\frac{g'}{g^3}-c\frac{f'}{fg^2}\right]'+\left(\frac{f'}{f}+3\frac{g'}{g}+\frac{1}{\rho}\right)\left[\left(d-2c\right)\frac{g'}{g^3}-c\frac{f'}{fg^2}\right]+\frac{1}{2}\mathfrak{T}=0,
\ee
\eseq
where $d=\frac{a+b}{2}$, and torsion scalar is
\beq\label{ts}
\mathfrak{T}=\left(c-d\right)\frac{f'^2}{f^2g^2}+(4c-2d)\frac{g'^2}{g^4}+4c\frac{f'g'}{fg^3},
\ee
and prime denotes differentiation with respect to $\rho$. Eqs. (\ref{te}) arise respectively from the $\mathfrak{G}_0^{\ 0}$, $\mathfrak{G}_1^{\ 1}$, and $\mathfrak{G}_2^{\ 2}=\mathfrak{G}_3^{\ 3}$ components of the Einstein equations. The temporal and angular equations can be rewritten in the form
\bseq\label{simpsys}
\beq\label{stet}
\left[\left(d-c\right)\frac{f'}{f}-2c\frac{g'}{g}\right]'+\left(\frac{f'}{f}+\frac{g'}{g}+\frac{2}{\rho}\right)\left[\left(d-c\right)\frac{f'}{f}-2c\frac{g'}{g}\right]+\frac{1}{2}\mathfrak{T}g^2=0,
\ee
\beq\label{stea}
\left[\left(d-2c\right)\frac{g'}{g}-c\frac{f'}{f}\right]'+\left(\frac{f'}{f}+\frac{g'}{g}+\frac{1}{\rho}\right)\left[\left(d-2c\right)\frac{g'}{g}-c\frac{f'}{f}\right]+\frac{1}{2}\mathfrak{T}g^2=0,
\ee
\eseq
and the radial equation, taking (\ref{ts}) into account, can be rewritten as
\beq\label{ster}
\frac{1}{2}\mathfrak{T}g^2=\frac{2}{\rho}\left[\left(d-2c\right)\frac{g'}{g}-c\frac{f'}{f}\right].
\ee
The first integrable equation is obtained by substituting (\ref{ster}) into (\ref{stea})
\beq\label{eq1}
\left[\left(d-2c\right)\frac{g'}{g}-c\frac{f'}{f}\right]'+\left(\frac{f'}{f}+\frac{g'}{g}+\frac{3}{\rho}\right)\left[\left(d-2c\right)\frac{g'}{g}-c\frac{f'}{f}\right]=0,
\ee
and the second integrable equation is obtained by subtracting twice (\ref{stea}) from (\ref{stet})
\beq\label{eq2}
\left[\left(c+d\right)\frac{g'}{g}+2(c-d)\frac{f'}{f}\right]'+\left(\frac{f'}{f}+\frac{g'}{g}+\frac{2}{\rho}\right)\left[\left(c+d\right)\frac{g'}{g}+2(c-d)\frac{f'}{f}\right]=0.
\ee
It turns out that the functions $f$ and $g$ from (\ref{sphtet}) choosing in the form of original article \cite{NewGR}
\beq\label{funcs}
f=\left(1-\frac{M}{s\rho}\right)^{s/2}\left(1+\frac{M}{q\rho}\right)^{-q/2},\quad g=\left(1-\frac{M}{s\rho}\right)^{1-s/2}\left(1+\frac{M}{q\rho}\right)^{1+q/2},
\ee
satisfy equations (\ref{eq1}) and (\ref{eq2}) when parameters $s, q, c, d$ satisfy the following algebraic equations 
\beq
qs(d-c)+(q-s)(2c-d)=0,\quad (3d-c)(q-s)+8(d-c)=0.
\ee
Substituting $s$, expressed from the second equation, into the first of these equations, we obtain an equivalent system
\beq
\left(\frac{c-d}{c-3d}\right)\cdot\bigg((3q^2+8q+8)d-(q+4)^2c\bigg)=0,\quad s=q+\frac{8(c-d)}{c-3d},
\ee 
from which two special cases arise
\bseq
\beq
c=d\quad \Longrightarrow\quad  s=q,
\ee
\beq
\frac{c}{d}=\frac{(3q^2+8q+8)}{(q+4)^2}\quad \Longrightarrow\quad q=\frac{2}{c-3d}\left(2d-2c\pm\sqrt{6dc-2d^2}\right).
\ee
\eseq
The first case called 1-Parameter New General Relativity and it was thoroughly considered in \cite{1pngr}. In the general second case using the substitution $c = d\left(\frac{2 + k^2}{6}\right)$, one can switch from the parameter pair $d, c$ to the pair $d, k$. And in this case, $q$ and $s$ have the form
\beq\label{sols}
q=-4\left(\frac{k\pm1}{k\pm4}\right),\quad s=4\left(\frac{k\mp1}{k\mp 4}\right).
\ee
It is important to note that the functions (\ref{funcs}) with coefficients (\ref{sols}) satisfy not only the system of two equations (\ref{simpsys}), but also the complete system (\ref{te}). This fact can be verified by direct substitution.

Thus, the exact static spherically symmetric vacuum solution of the NGR field equations is a tetrad field of the form (\ref{sphtet}), which induces the metric with restored dimensionality
\begin{multline}\label{exs}
ds^2=\mathfrak{c}^2\left(1-\frac{M(k-4)}{4(k-1)\rho}\right)^{\frac{4(k-1)}{k-4}}\left(1-\frac{M(k+4)}{4(k+1)\rho}\right)^{\frac{4(k+1)}{k+4}}dt^2-\\-\left(1-\frac{M(k-4)}{4(k-1)\rho}\right)^{\frac{4+2k}{4-k}}\left(1-\frac{M(k+4)}{4(k+1)\rho}\right)^{\frac{4-2k}{4+k}}\left(d\rho^2+\rho^2d\theta^2+\rho^2\sin^2\theta d\phi^2\right),
\end{multline}
where $k=\pm\sqrt{12c/(a+b)-2}$, $\mathfrak{c}$ is the speed of light, $M=mG/\mathfrak{c}^2$, $G$ is Newtonian gravitational constant. It is important to note that both solutions (\ref{sols}) corresponding to different signs give the same metric (\ref{exs}). In original work \cite{NewGR} and in the paper \cite{sisngr} this solution was given in a slightly different and more cumbersome parametrization.

At $k=\pm2$ this metric, as expected, transforms into the Schwarzschild \cite{bh1916} metric in isotropic coordinates
\beq\label{schwarz}
ds^2=\mathfrak{c}^2\left(\frac{1-\frac{M}{2\rho}}{1+\frac{M}{2\rho}}\right)^2dt^2-\left(1+\frac{M}{2\rho}\right)^4\left(d\rho^2+\rho^2d\theta^2+\rho^2\sin^2\theta d\phi^2\right).
\ee
\section{About exact solution}
It may seem that solution (\ref{exs}) has two zeros
\beq
\rho_1=\frac{M(k-4)}{4(k-1)},\quad \rho_2=\frac{M(k+4)}{4(k+1)},
\ee
However, it is easy to see that both zeros are positive for $k\in(-\infty;-4)\cup(-1;1)\cup(4;\infty)$, and these regions do not contain cases $k=\pm2$, which realise TEGR. That is, two horizons can appear in models whose physical properties differ greatly from those of GR. Nevertheless, for a more careful analysis, it is useful to construct the curvature scalar
\beq
R=\frac{3M^2k^2(4-k^2)(M-4\rho)^2}{32\rho^6(k^2-1)^2\left(1-\frac{M(k-4)}{4(k-1)\rho}\right)^{\frac{12}{4-k}}\left(1-\frac{M(k+4)}{4(k+1)\rho}\right)^{\frac{12}{4+k}}},
\ee
which, in the vicinity of $k=\pm2$, to leading order in smallness has the form 
\beq
R=\frac{128M^2(M-4\rho)^2\rho^2}{3(M+2\rho)^6(M-2\rho)^2}(2\mp k)+O\bigg((2\mp k)^2\bigg),
\ee
from which it is clear that the curvature scalar is singular at $\rho=M/2$ in all models except $k=\pm2$. Rich phenomenology of these models, as we mentioned earlier, was either described in the work \cite{1pngr} or reduced to TEGR, therefore, our further analysis will be devoted to the study of classical effects in models other than 1-Parameter New General Relativity. 
\section{Tests}
 
To make a quantitative test of Einstein's theory of relativity we should study geodesics in a specific gravitational field. Therefore, before describing the observable effects in NGR, we will write out the geodesic equations for the found metric.
The geodesic equations in spacetime (\ref{exs}) have the form
\beq
\theta=\frac{\pi}{2},\quad 
\dot \phi=\frac{L}{g^2\rho^2},\quad
\dot t=\frac{E}{\mathfrak{c}^2f^2},\quad
\mathfrak{c}^2f^2\dot t^2-g^2\left(\dot \rho^2+\rho^2\dot \theta^2 +\rho^2\sin^2\theta \dot\phi^2\right)=\delta \mathfrak{c}^2,
\ee
where $E, L$ are integrals of motion, $\delta=1$ for timelike geodesics and $\delta=0$ for null geodesics and dot denotes the derivative with respect to a parameter along the curve - proper time $\tau$. Eliminating $\dot t$ and $\dot\phi$ from the fourth equation, we obtain a differential equation
\beq
\frac{E^2}{\mathfrak{c}^2f^2g^2} -\frac{L^2}{g^4\rho^2}-\delta \frac{\mathfrak{c}^2}{g^2}=\dot \rho^2,
\ee
for $\rho(\tau)$ in which the variables can be separated. On the right-hand side we can change the differentiation with respect to $\tau$ to differentiation with respect to $\phi$ using $\frac{d\rho}{d\tau}=\frac{d\rho}{d\phi}\frac{d\phi}{d\tau}=\frac{d\rho}{d\phi}\cdot\frac{L}{g^2\rho^2}$ and reduce the equation to
\beq
\rho'^2+\rho^2=\rho^4\left[\left(\frac{Eg}{L\mathfrak{c}f}\right)^2-\delta\left(\frac{\mathfrak{c}g}{L}\right)^2\right].
\ee
Binet’s substitution $\rho = \frac{1}{u}$ takes the form
\beq
(u')^2+u^2=\left(\frac{Eg}{L\mathfrak{c}f}\right)^2-\delta\left(\frac{\mathfrak{c}g}{L}\right)^2.
\ee
Differentiating this equation with respect to $\phi$ we obtain
\beq
u''+u=\frac{1}{2L^2u^2}\left.\left(\delta\mathfrak{c}^2\frac{d}{d\rho}(g^2)-\left(\frac{E}{\mathfrak{c}}\right)^2\frac{d}{d\rho}\left(\frac{g^2}{f^2}\right)\right)\right|_{\rho=u^{-1}}.
\ee
Due to the complexity of the functions $f^2$ and $g^2$ in the definition of the metric (\ref{exs}), we will only expand the right-hand side into a series up to the first nonlinear order in $u$ around zero. Then the equation takes the form

\beq\label{beq}
u''+u=\frac{3ME^2}{2L^2\mathfrak{c}^2}\left(\frac{k^2}{k^2-1}+\frac{Mk^2(13k^2-7)u}{4(k^2-1)^2}\right)-\frac{\delta\mathfrak{c}^2M}{2L^2}\left(\frac{k^2+2}{k^2-1}+\frac{M(5k^4+5k^2+8)u}{4(k^2-1)^2}\right)+O(u^2).
\ee
This equation at $k=2$ becomes a relativistic generalization of the mechanical Binet equation in the Schwarzschild spacetime, which can be used to describe the anomalous perihelion shift, time delay (for $\delta=1$) and the deflection of light rays (for $\delta=0$). They will be needed to constrain the parameter $k$.
\subsection{Precession of planetary orbits}
For the motion of massive bodies with $\delta=1$ the linear order of Eq. (\ref{beq}) is
\beq
u''+u=\frac{3ME^2}{2L^2\mathfrak{c}^2}\left(\frac{k^2}{k^2-1}+\frac{Mk^2(13k^2-7)u}{4(k^2-1)^2}\right)-\frac{M\mathfrak{c}^2}{2L^2}\left(\frac{k^2+2}{k^2-1}+\frac{M(5k^4+5k^2+8)u}{4(k^2-1)^2}\right),
\ee
which is an ordinary harmonic oscillator equation with a constant external force $u''+(1-\be)u=\al$, where
\beq
\al=\frac{M\mathfrak{c}^2}{2L^2(k^2-1)}\left(3k^2(E/\mathfrak{c}^2)^2-k^2-2\right),\quad 
\be=\frac{M^2\mathfrak{c}^2}{8L^2(k^2-1)^2}\left(3k^2(E/\mathfrak{c}^2)^2(13k^2-7)-5k^4-5k^2-8\right).
\ee
Its solution is the function $u = \al + \frac{\cos\left(\sqrt{1-\be}\phi\right)}{\rho_0} = \frac{1}{\rho}$. Maximum of $u(\phi)$, i.e. the minimum of $\rho(\phi)$ is firstly reached at $\phi = 0$, and then at $\phi = \frac{2\pi}{\sqrt{1-\be}} \approx 2\pi + \pi\be$, where $\pi\be$ is the perihelion shift.

\begin{figure}
\begin{center}
\begin{tikzpicture}
\begin{polaraxis}[
    width=6cm,
    height=6cm,
    hide axis,
    xmin=0, xmax=360,
    ymin=0, ymax=1.5,
    clip=false
]
\addplot[domain=0:360, samples=181, thick, black]
    {0.8/(1+0.7*cos(x))};
\addplot[domain=0:360, samples=181, thick, black, dashed]
    {0.8/(1+0.7*cos(0.97*x+15))};

\fill[black] (0,0) circle (1pt);
\node[below right] at (-1,0) {\footnotesize $M$};

\coordinate (P1) at (0, {0.8/(1+0.7)});
\coordinate (P2) at ({15/0.97}, {0.8/(1+0.7)});

\draw[gray, thick] (0,0) -- (P1);
\draw[gray, thick] (0,0) -- (P2);

\node at (0,0.8) {\footnotesize $\Delta\phi$};

\fill (P1) circle (1.5pt);
\fill (P2) circle (1.5pt);

\end{polaraxis}
\end{tikzpicture}
\caption{Precession of perihelion}
\label{pp}
\end{center}
\end{figure}
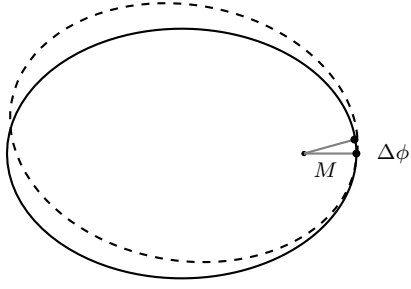

\beq
\Delta\phi=\pi\be=\frac{\pi m^2G^2}{8\mathfrak{c}^2L^2}\cdot\frac{\left(3k^2(E/\mathfrak{c}^2)^2(13k^2-7)-5k^4-5k^2-8\right)}{(k^2-1)^2}\approx\frac{3\pi m^2G^2}{2\mathfrak{c}^2L^2}\cdot\bigg[\left(5(E/\mathfrak{c}^2)^2-1\right)-\left(\frac{23}{9}(E/\mathfrak{c}^2)^2-1\right)(k-2)\bigg],
\ee
where we used the series expansion of a function of $k$ around the point $k=2$. Assuming the motion of the test body non-relativistic, $E/\mathfrak{c}^2 \approx 1$, we have
\beq\label{r1}
\Delta\phi_{\text{\tiny{NGR}}}=\frac{6\pi m^2G^2}{\mathfrak{c}^2L^2}\left(1-\frac{7}{18}(k-2)\right)=\Delta\phi_{\text{\tiny{TEGR}}}\left(1-\frac{7}{18}(k-2)\right).
\ee
Equation (\ref{r1}) can give us values of $k$ which do not lead to contradiction with NGR and experiments.
\subsection{Deflection of light rays}
The propagation of light is described by null geodesics, in this case the parameter $\delta$ is set to zero, so Eq. (\ref{beq}) has the form
\beq
u''+u=\frac{3ME^2}{2L^2\mathfrak{c}^2}\left(\frac{k^2}{k^2-1}+\frac{Mk^2(13k^2-7)u}{4(k^2-1)^2}\right),
\ee

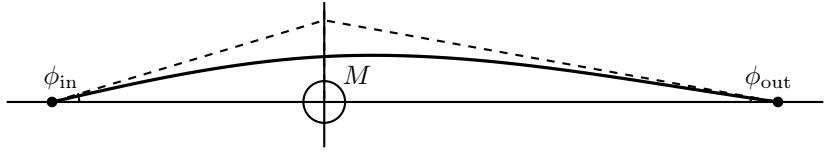
\begin{figure}[htbp]
\centering
\begin{tikzpicture}[scale=1.2]
    \draw[thick] (-3.5,0) -- (5.5,0);
    \draw[thick] (0,-0.5) -- (0,1.1);
     
    \node[above right] at (0.1,0.1) {$M$};
    
    \draw[dashed, thick, black] (-3,0) -- (0,0.9) -- (5,0);
    \draw[dashed, thick, black] (0,1) -- (0,0);
    
    \draw[very thick, black, smooth] plot[domain=-3:0, samples=50] 
        (\x, {1/2 + 4*\x/75 - 4*\x*\x/75 - 7*\x*\x*\x/1350});
    
    \draw[very thick, black, smooth] plot[domain=0:5, samples=50] 
        (\x, {1/2 + 4*\x/75 - 4*\x*\x/75 + 17*\x*\x*\x/3750});
    
    \draw[black, thick] (0,0) circle (0.23);
    
    \filldraw[black] (-3,0) circle (1.5pt);
    \filldraw[black] (5,0) circle (1.5pt);
    
    \draw[thick] (-3,0) ++(0.3,0) arc (0:18.435:0.3);
    \node at (-2.9,0.27) {$\phi_{\text{in}}$};
    
    \draw[thick] (5,0) ++(-0.3,0) arc (180:168.69:0.3);
    \node at (4.9,0.27) {$\phi_{\text{out}}$};
    
\end{tikzpicture}
\caption{Bending of light beam}
\label{gcf}
\end{figure}

It turns out that to describe the effect of gravitational light deflection, the linear term in $M$ on the right-hand side is sufficient. And equation $u''+(1-\be)u=\al$ with parameters
\beq
\al=\frac{3ME^2}{2L^2\mathfrak{c}^2}\cdot\frac{k^2}{k^2-1},\quad \be=\frac{3M^2E^2}{8L^2\mathfrak{c}^2}\cdot\frac{k^2(13k^2-7)}{(k^2-1)^2},
\ee
is solved by the function
\beq
u=\al+\frac{\sin(\sqrt{1-\beta}\phi)}{\rho_0},
\ee
where $\rho_0$ is the impact parameter, and the second integration constant $\phi_0$ (initial phase) is omitted for simplicity of the expressions. $u$ vanishes for small $\phi$ and for $\phi$ close to $\pi$. Therefore, expanding $\sin\phi$ around the points of interest, we obtain two equations
\beq
0=\al+\frac{\sqrt{1-\beta}\phi_{\text{\tiny{in}}}}{\rho_0},\quad 0=\al+\frac{(\pi-\sqrt{1-\beta}\phi_{\text{\tiny{out}}})}{\rho_0},
\ee
so 
\beq\label{dphi}
\phi_{\text{\tiny{out}}}-\phi_{\text{\tiny{in}}}=\Delta\phi=\frac{\pi+2\al\rho_0}{\sqrt{1-\beta}}=\pi+\frac{3ME^2}{L^2\mathfrak{c}^2}\cdot\frac{k^2}{k^2-1}+O(M^2).
\ee
In the case of flat spacetime the difference between the exit and entry angles of the light beam is $\phi_{\text{\tiny{out}}}-\phi_{\text{\tiny{in}}}=\pi$. Therefore, the gravitational contribution is determined only by the second term. The energy and angular momentum can be related to the impact parameter by the relation $L = (E/\mathfrak{c})\rho_0$. Hence, expanding the function of $k$ in equation \eqref{dphi} around $k = 2$ one obtains the expression
\beq\label{r2}
\delta\phi_{\text{\tiny{NGR}}}=\frac{4mG}{\rho_0\mathfrak{c}^2}\left(1-\frac{k-2}{3}\right)=\delta\phi_{\text{\tiny{TEGR}}}\left(1-\frac{k-2}{3}\right),
\ee
that can be used to estimate the value of $k$ at which the predictions of NGR would not contradict experiment.
\subsection{Shapiro time delay}
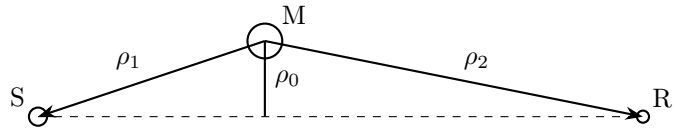
\begin{figure}[h]
    \centering
    \begin{tikzpicture}[>=Stealth, scale=1]

        \coordinate (O1) at (0,0);
        \coordinate (O2) at (-3,-1);
        \coordinate (O3) at (5,-1);
        
        \draw[thick] (O1) circle (0.23);
        \draw[thick] (O2) circle (0.12);
        \draw[thick] (O3) circle (0.08);
        
        \node[above right] at (0.1,0.1) {M};
        \node[below left] at (-3.05,-0.5) {S};
        \node[above right] at (5,-1) {R};
       
        \draw[->, thick] (O1) -- node[midway, above left] {$\rho_1$} (O2);
        \draw[->, thick] (O1) -- node[midway, above right] {$\rho_2$} (O3);
        
        \draw[dashed] (O2) -- (O3);
        \coordinate (H) at ($(O2)!(O1)!(O3)$);
        \draw[ thick] (O1) -- node[midway, right] {$\rho_0$} (H);
        
    \end{tikzpicture}
\caption{Diagram of the arrangement of light source S, the receiver R, and the gravitating mass M.}
\label{shef}
\end{figure}

To describe the Shapiro effect, it is necessary to consider the direct and reverse passage of light between the source and the receiver using geodesic equations
\beq\label{gesh}
\dot \phi=\frac{L}{g^2\rho^2},\quad
\dot t=\frac{E}{\mathfrak{c}^2f^2},\quad
\mathfrak{c}^2f^2\dot t^2-g^2\left(\dot \rho^2 +\rho^2\dot\phi^2\right)=0.
\ee
The diagram of the light's path from the source to the receiver and back is shown in Fig. \ref{shef}. 

In flat space the distance traveled by a light and the time of motion are related by the expression 
\beq\label{notrel}
\mathfrak{c}\Delta t_{\text{not rel}}=2\left(\sqrt{\rho_1^2-\rho_0^2}+\sqrt{\rho_2^2-\rho_0^2}\right).
\ee
At the closest point $\rho_0$, function $\rho(\tau)$ reaches its minimum, and the Eqs. (\ref{gesh}) take the form
\beq\label{min}
\dot \phi_0=\frac{L}{g_0^2\rho_0^2},\quad
\dot t_0=\frac{E}{\mathfrak{c}^2f_0^2},\quad
\mathfrak{c}^2f_0^2\dot t_0^2-g_0^2\rho_0^2\dot\phi_0^2=0,
\ee
where all quantities with a zero subscript correspond to the moment of proper time when $\rho=\rho_0$. The set of equations (\ref{min}) allows us to express $L$ in terms of $\rho_0$ in the form $L= \rho_0 (E/\mathfrak{c})(g_0/f_0)$, which makes it possible to rewrite the third of Eqs. (\ref{gesh}) as
\beq
1 -\left(\frac{\rho_0}{\rho}\right)^2\left(\frac{g_0}{f_0}\right)^2\left(\frac{f}{g}\right)^2=\left(\frac{g}{f}\right)^2\left(\frac{d\rho}{d t}\right)^2\frac{1}{\mathfrak{c}^2},
\ee
where the variables can be separated
\beq
\mathfrak{c}dt=\frac{d\rho}{\sqrt{(f/g)^2 -(\rho_0/\rho)^2(g_0/f_0)^2\left(f/g\right)^4}}.
\ee
The right-hand side of this expression after integration cannot be expressed in terms of elementary functions, so we consider just expansion of the integrand into a series for small $M$. In other words,  full time of motion in the weak-field limit is
\beq
\mathfrak{c}\Delta t=2\left(F(\rho_0,\rho_1)+F(\rho_0,\rho_2)\right),
\ee
where 
\beq
F(\rho_0,\rho)=\sqrt{\rho^2-\rho_0^2}+\frac{3M}{2}\left(\ln\left(\rho/\rho_0+\sqrt{(\rho/\rho_0)^2-1}\right)+\sqrt{\frac{\rho-\rho_0}{\rho+\rho_0}}\right)\cdot\frac{k^2}{k^2-1},
\ee
therefore, the relativistic expression for the time taken for a signal to travel from the source to the receiver and back in the leading order for $\rho_1$ and $\rho_2$ takes the form
\beq
\mathfrak{c}\Delta t_{\text{rel}}=\mathfrak{c}\Delta t_{\text{not rel}}+\frac{3mG}{\mathfrak{c}^2}\ln\left(4\rho_1\rho_2/\rho_0^2\right)\cdot\frac{k^2}{k^2-1}.
\ee
Here the first term was defined in (\ref{notrel}), but the second term is Shapiro delay
\beq\label{r3}
\mathfrak{c}\Delta t_{\text{S \tiny{NGR}}}=\frac{4mG}{\mathfrak{c}^2}\ln\left(4\rho_1\rho_2/\rho_0^2\right)\left(1-\frac{k-2}{3}\right)\equiv\mathfrak{c}\Delta t_{\text{S \tiny{TEGR}}}\left(1-\frac{k-2}{3}\right),
\ee
where the series expansion around $k=2$ was used. It should be noted that the relative contribution of parameter $k$ to this effect is the same as in the light deflection (\ref{r2}). It is obvious since both effects arise from solving the equations for lightlike geodesics. And from the resulting expression (\ref{r3}) for the time delay, one can obtain experimental constraints on $k$.
\subsection{de Sitter geodetic precession}
The spin (angular momentum) vector $S^\al$ of a gyroscope moving along a geodesic satisfies the covariant constancy equation
\beq
\frac{D S^\al}{d\tau}=0,
\ee
where $\tau$ is proper time. This equation in components reads
\beq
u^\mu\nabla_\mu S^\al=0,\qquad \text{or}\qquad \frac{d S^\al}{d\tau}+u^\mu\Gamma^\al_{\mu\nu}S^\nu=0,
\ee
which for metric (\ref{exs}) and in equatorial plane of free movement $\theta=\pi/2$ for the circle trajectory $\rho=\rho_0=\text{const}$ constitutes a system of ordinary differential equations
\beq
\dot{S}^0 + \frac{f'}{f}\dot{t}\,S^1  = 0,\quad
\dot{S}^1 + \frac{\mathfrak{c}^2 f f'}{g^2}\,\dot{t}\,S^0  - \rho_0^2\Bigl(\frac{g'}{g}+\frac{1}{\rho_0}\Bigr)\dot{\phi}\,S^3= 0,\quad
S^2=\text{const},\quad 
\dot{S}^3 + \Bigl(\frac{g'}{g} + \frac{1}{\rho_0}\Bigr) \dot{\phi}\,S^1 = 0,
\ee
where $\frac{f'}{f}$, $\frac{g'}{g}$, $\dot{t}$, $\dot{\phi}$ are constants defined by $\rho=\rho_0$. After ignoring $S^2$ and eliminating $S^0$, the system can be reduced to a pair of independent equations
\beq
\ddot{S}^a+\omega^2S^a=C^3\delta^a_3,
\ee
where $a=1,3$, $C^3$ - is integration constant, and the square of the oscillation frequency
\beq
\omega^2=\left(\frac{g'}{g}+\frac{1}{\rho}\right)^2\frac{L^2}{\rho^2g^4}-\frac{E^2}{\mathfrak{c}^2f^4}\left(\frac{f'}{g}\right)^2\bigg|_{\rho=\rho_0},
\ee
the difference of which with $\dot\phi$ determines de Sitter precession rate $\Omega=\dot\phi-\omega$, where $L$ can be expressed in terms of $\rho_0$ from the circular orbit stability condition $\ddot\rho=0$
\beq
L^2=\left(\mathfrak{c}\rho\right)^2\cdot\frac{\left(\frac{E}{\mathfrak{c}^2}\right)^2\left(\frac{g}{f}\right)^2\left(\frac{f'}{f}+\frac{g'}{g}\right)-gg'}{2\frac{g'}{g}+\frac{1}{\rho}}\bigg|_{\rho=\rho_0}.
\ee
Thus, in the weak field limit $\rho_0\to\infty$, in the vicinity of $k=2$, the geodesic precession angular frequency is
\beq\label{r4}
\Omega_{\text{\tiny{NGR}}}=\frac{3mG}{2\mathfrak{c}^2\rho_0}\omega_{\text{\tiny{N}}}\left(1-\frac{4}{9}\left(k-2\right)\right)=\Omega_{\text{\tiny{TEGR}}}\left(1-\frac{4}{9}\left(k-2\right)\right),
\ee
where $\omega_{\text{\tiny{N}}}=\sqrt{mG/\rho_0^3}$ - the Newtonian circular frequency, and dimensionless mass was replaced by the dimensional one $M=mG/\mathfrak{c}^2$, and since the motion of any massive body along this circular orbit is non‑relativistic, the approximation $E/\mathfrak{c}^2\approx1$ was used. And from resulting expression (\ref{r4}) for the angular frequency, and hence for the angular precession, one can obtain experimental constraints on the value of $k$. 
\subsection{Red shift}
Using the well-known redshift relation for metric (\ref{exs})
\beq
1+z=\frac{f(\rho_2)}{f(\rho_1)},
\ee
expanding the right-hand side into a series in the variables $G/\mathfrak{c}^2$ and $(k-2)$ around zero, one obtains
\beq\label{grshk}
z=\frac{mG}{\mathfrak{c}^2}\left(\frac{1}{\rho_1}-\frac{1}{\rho_2}\right)+\frac{m^2G^2}{2\mathfrak{c}^4}\left(\frac{1}{\rho_1}-\frac{1}{\rho_2}\right)^2+\frac{m^2G^2}{6\mathfrak{c}^4}\left(\frac{1}{\rho_1^2}-\frac{1}{\rho_2^2}\right)(k-2)+O\left(\left(\frac{mG}{\mathfrak{c}^2}\right)^3\right).
\ee
The first two terms determine the magnitude of the redshift in TEGR in the weak field limit up to the second order of smallness, the term proportional to $(k-2)$ is of the third order of smallness. Therefore, neglecting higher-order terms and assuming that $\rho_2\gg\rho_1$ we have
\beq\label{rrsh}
z_{\text{\tiny{NGR}}}=z_{\text{\tiny{TEGR}}}\left[1+\frac{z_{\text{\tiny{TEGR}}}\left(k-2\right)}{6}\right].
\ee
The resulting expression (\ref{rrsh}) can also be used to obtain experimental constraints on the parameter $k$. However, the smallness $z^2_{\text{\tiny{TEGR}}}$ of will strongly suppress the factor $(k-2)$, as will be shown below.
\section{Summary}
\subsection{Comparison and assessment}
In modern experimental reviews, values are typically given not for the directly measured quantities — such as the perihelion shift, light deflection, time delay, or redshift — but for the post-Newtonian parameters \cite{ppn} $\beta,\gamma$.
PPN expansion of the static spherically symmetric metric. in the weak-field limit is
\beq\label{ppnexpan}
g_{00}/\mathfrak{c}^2=1-\frac{2M}{\rho}+2\beta\frac{M^2}{\rho^2}+O\left(\frac{M^3}{\rho^3}\right), \quad g_{ij}=-\left(1+2\gamma\frac{M}{\rho}\right)\delta_{ij}+O\left(\frac{M^2}{\rho^2}\right).
\ee 
In terms of these parameters, all effects are represented as:
\bseq \label{exper}
\beq \label{expera}
\Delta \phi=\frac{6\pi m^2G^2}{\mathfrak{c}^2L^2}\left(\frac{2+2\ga-\be}{3}\right),
\ee
\beq\label{experb}
\delta\phi=\frac{4mG}{\rho_0\mathfrak{c}^2}\left(\frac{1+\ga}{2}\right), 
\ee
\beq\label{experc}
\mathfrak{c}\Delta t_{\text{S}}=\frac{4mG}{\mathfrak{c}^2}\ln\left(4\rho_1\rho_2/\rho_0^2\right)\left(\frac{1+\ga}{2}\right),
\ee
\beq\label{expere}
\Omega=\frac{3mG}{2\mathfrak{c}^2\rho_0}\omega_{\text{\tiny{N}}}\left(\frac{1+2\gamma}{3}\right).
\ee
\beq\label{experd}
z=\frac{mG}{\mathfrak{c}^2}\left(\frac{1}{\rho_1}-\frac{1}{\rho_2}\right)+\frac{m^2G^2}{2\mathfrak{c}^4}\left(\frac{1}{\rho_1}-\frac{1}{\rho_2}\right)^2-(\beta-1)\frac{m^2G^2}{\mathfrak{c}^4}\left(\frac{1}{\rho_1^2}-\frac{1}{\rho_2^2}\right),
\ee
\eseq

Thus, for the estimation of the parameter $k$ via the perihelion shift, the results of \cite{coex} will be used:  
\beq\label{ppnp}
\be-1=(-4.5\pm 5.6)\times 10^{-5},\quad \ga-1=(-1.2\pm 1.2)\times 10^{-5}.
\ee 
To estimate the value of $k$ from light deflection and time delay, one can use either $\gamma-1$ from \cite{coex} or, for example, the Cassini data \cite{cas}:
\beq\label{expcas}
\gamma-1=(2.1\pm 2.3)\times 10^{-5}.
\ee
Taking into account the connection of parameters (\ref{ppnp}) or (\ref{expcas}) with the quantities under study (\ref{exper}), it is not difficult to estimate the value of the parameter $k$ from the expressions (\ref{r1}), (\ref{r2}), (\ref{r3}), (\ref{r4}). 
\begin{enumerate}
\item From the first expression (\ref{r1}) and  (\ref{expera}) we have
\beq
\frac{2+2\ga-\be}{3}=1-\frac{7}{18}(k-2),
\ee
it leads to $k-2=(1.8\pm 5.2)\times 10^{-5}$ from (\ref{ppnp}) .
\item From the second expression (\ref{r2}) (or third (\ref{r3})) and (\ref{experb}) (or third (\ref{experc})) it is easy to obtain
\beq
\frac{1+\gamma}{2}=1-\frac{k-2}{3},
\ee 
it leads to the first estimate of the parameter value $k-2=(1.8 \pm 1.8)\times 10^{-5}$ from (\ref{ppnp}) and  $k-2=(-3.2 \pm 3.5)\times 10^{-5}$ from (\ref{expcas}).
\item From the fourth expression (\ref{r4}) and  (\ref{expere}) we have
\beq
\frac{1+2\gamma}{3}=1-\frac{4}{9}\left(k-2\right).
\ee
it leads to $k-2=(1.8 \pm 1.8)\times 10^{-5}$ from (\ref{ppnp}) and $k-2=(-3.2 \pm 3.5)\times 10^{-5}$ from (\ref{expcas}), which matches the previous point, since in both cases $k-2=\frac{3}{2}(1-\gamma)$.
\item  After comparing formulas (\ref{experd}) and (\ref{grshk}), it may seem that the value of the parameter $k$ can be easily estimated using $\beta$, simply by equating the coefficients of $\left(mG/\mathfrak{c}^2\right)^2$ and obtaining $6(1-\beta)=k-2$. However, this is not the case, because even the most modern experimental tests of gravitational redshift \cite{26grsh} use only the first approximation proportional to potential difference $\Delta U$. Whereas $\beta$ in (\ref{experd}) and $k$ in (\ref{grshk}) appear only in quadratic terms of the gravitational potential.
Therefore it is not effective to use gravitational redshift to estimate parameter $k$. This fact can be demonstrated using classical results of the Pound–Rebka experiment \cite{PR} testing general relativity. Where theoretical value is $z_{\text{\tiny{GR}}}=z_{\text{\tiny{TEGR}}}= 2.46 \times 10^{-15}$ and the experimental estimation is $z = (2.57 \pm 0.26) \times 10^{-15}$. Such experimental value leads to $k-2=(1.09\pm2.58)\times 10^{14}$ in NGR. The reason for such a weak constraint on the parameter $k$ is that in formula (\ref{rrsh}) the quantity $k-2$ is suppressed by the second order of smallness of $mG/\mathfrak{c}^2$. This arises because the metric  function $f^2$ (\ref{exs}) in the weak-field limit has the form
\beq
f^2=1-\frac{2M}{\rho}+\frac{(7k^2-4)M^2}{4(k^2-1)\rho^2}+O(M^3),
\ee
which, to leading order, coincides with $g_{00}$ in (\ref{ppnexpan}). Gravitational redshift does not imply significant constraints on $k$ due to absence of $k$ in the first order of both expansions. 
\end{enumerate}

\subsection{Conclusion}
In this work, an exact static spherically symmetric vacuum solution of the New General Relativity field equations has been obtained. It has been shown that the metric (\ref{exs}) contains, in addition to the mass $M$, a dimensionless parameter $k$. For $k = \pm 2$, the solution reduces to the classical Schwarzschild metric in isotropic coordinates (\ref{schwarz}), which corresponds to General Relativity and its teleparallel equivalent (TEGR). Thus, NGR predicts a one-parameter family of gravitational fields outside spherical gravitating bodies, which coincides with the Schwarzschild field to leading order but differs at higher orders.

Using the four classical gravitational tests (perihelion precession, light deflection, Shapiro time delay, de Sitter geodetic precession and gravitational redshift), expressions for the observable effects have been derived to the first nonlinear order in $M$. It is shown that all effects, except for the redshift, depend linearly on the deviation of the parameter $k$ from the value $2$. Comparing these expressions with modern experimental constraints on the PPN parameters $\beta$ and $\gamma$ yields the following estimates:

\begin{itemize}
    \item From perihelion precession (lunar laser ranging data \cite{coex}):
    $$
    k - 2 = (1.8 \pm 5.2) \times 10^{-5}.
    $$
    \item From light deflection, time delay and geodetic precession (same data \cite{coex}):
    $$
    k - 2 = (1.8 \pm 1.8) \times 10^{-5},
    $$
    and from the Cassini mission experiment \cite{cas}:
    $$
    k - 2 = (-3.2 \pm 3.5) \times 10^{-5}.
    $$ 
    \item Gravitational redshift (Pound--Rebka experiment \cite{PR}) yields only a very weak constraint $k - 2 \sim 10^{14}$ because the parameter $k-2$ appears only in terms of order $(mG/\mathfrak{c}^2)^2$ and higher.
\end{itemize}

Thus, within current experimental accuracy, NGR is not in conflict with observations, and the most precise estimate of the parameter $k$ that we have obtained is $k-2 = (1.8 \pm 1.8)\times 10^{-5}$, based on data from \cite{coex}. Further tests of NGR and the magnitude of its parameters can be carried out using effects in binary pulsars, black hole shadows or their spectroscopy; cosmological predictions of NGR can also be considered. 
\section*{Declarations}
\begin{itemize}
\item \textbf{Funding:} No funding was received for conducting this study or preparing the manuscript. 
\item \textbf{Competing interests:} The authors declare no competing interests. 
\item \textbf{Availability of data and materials:} Not applicable.
\end{itemize}

\end{document}